\documentclass[a4paper, 11pt]{article}

\usepackage{fullpage}
\usepackage{amsmath,amssymb,float,graphicx,multirow,enumerate}
\usepackage{hyperref}
\hypersetup{colorlinks = true, linktocpage = true, bookmarksopen = true, linkcolor = blue, urlcolor=blue, citecolor = blue, allcolors=blue}
\usepackage[caption=false]{subfig}
\usepackage[font=footnotesize]{caption}
\usepackage[numbers,sort&compress]{natbib}

\usepackage[table]{xcolor}
\def\cbl{\color{black}}
\def\cb{\color{black}}
\newcommand\change[1]{{\color{black}{#1}}}

\begin{document}

%\title{Characterising the timescale for diffusion through layers and across interfaces}
\title{Accurate and efficient calculation of \change{response times}  for groundwater flow}

\author{Elliot\hspace{0.1cm}J.\hspace{0.1cm}Carr\footnote{Corresponding author: \href{mailto:elliot.carr@qut.edu.au}{elliot.carr@qut.edu.au}}\hspace*{0.15cm} and Matthew J. Simpson\\
\small School of Mathematical Sciences, Queensland University of Technology (QUT), Brisbane, Australia.}
\date{}
%\date{\small 16 January 2017}
%\email[]{elliot.carr@qut.edu.au}
%\homepage[]{Your web page}
%\thanks{}
%\altaffiliation{}
%\affiliation{School of Mathematical Sciences, Queensland University of Technology (QUT), Brisbane, Australia}

\maketitle

\begin{abstract}
We study measures of the amount of time required for transient flow in heterogeneous porous media to effectively reach steady state, also known as the response time.  Here, we develop a new approach that extends the concept of mean action time.  Previous applications of the theory of mean action time to estimate the response time use the first two central moments of the probability density function associated with the transition from the initial condition, at $t=0$, to the steady state condition that arises in the long time limit, as $t \to \infty$.  This previous approach leads to a computationally convenient estimation of the response time, but the accuracy can be poor.  Here, we outline a powerful extension using the first $k$ raw moments, showing how to produce an extremely accurate estimate by making use of asymptotic properties of the cumulative distribution function.  Results are validated using an existing laboratory-scale data set describing flow in a homogeneous porous medium.  In addition, we demonstrate how the results also apply to flow in heterogeneous porous media.  Overall, the new method is: (i) extremely accurate; and (ii) computationally inexpensive.  In fact, the computational cost of the new method is orders of magnitude less than the computational effort required to study the response time by solving the transient flow equation.  Furthermore, the approach provides a rigorous mathematical connection with the heuristic argument that the response time for flow in a homogeneous porous medium is proportional to $L^2/D$, where $L$ is a relevant length scale, and $D$ is the aquifer diffusivity.  Here, we extend such  heuristic arguments by providing a clear mathematical definition of the proportionality constant.
\end{abstract}

%\begin{keyword}
%multilayer, diffusion, timescale, mean action time, transition time.
%\end{keyword}

%\pacs{}

%\maketitle

\section{Introduction}

Transient, or time dependent groundwater flow conditions are more complicated than steady state groundwater flow conditions~\citep{Bear1972,Bear1979}.  The physical differences in complexity are echoed in the differences between mathematical models of transient groundwater flow and mathematical models of steady state groundwater flow, with the latter simpler to solve than the former~\citep{anderson_2007,Haitjema1995,Wang1983}.  This is because steady state groundwater flow models are elliptic partial differential equations that do not involve the specification of the initial condition or storage parameter.  In contrast, transient groundwater flow models are parabolic partial differential equations that require the specification of both the initial condition and the storage parameter associated with the porous material.   Since steady state flow conditions arise as the long time limit of a transient flow response~\citep{Haitjema1995}, is it natural for us to determine an estimate of the amount of time required for a transient response to occur, after which steady state conditions will prevail and simpler steady models can be used to describe the flow process.  Such a time scale is often referred to as a \textit{response time}~\citep{Bredehoeft_2009,currell2016,haitjema2006}.

In the groundwater modelling literature there are two main techniques used to calculate the response time.  In the first approach, both the transient groundwater flow model and the steady state groundwater flow models are solved, and the response time is taken to be the amount of time taken for the difference between the transient solution and the associated steady state solution to fall below some sufficiently small tolerance~\citep{RousseauGueutin_2009,lu_2013,watson_2010}.  In the second method a simple scaling approach is adopted whereby if groundwater flow takes place in a confined aquifer with aquifer diffusivity $D$, then the response time is proportional to $L^2/D$, where $L$ is a relevant length scale~\citep{Bredehoeft_2009,currell2016,haitjema2006}. Both of these methods suffer from certain limitations.  For example, the first method relies on solving both the steady state and transient flow problem of interest.  We note that if the transient solution is used to study the response time then this necessarily involves studying the long time limit, $t\rightarrow\infty$.  Furthermore, in a standard implicit scheme, small time steps are required to control truncation error.  Therefore, these two requirements, combined, mean that the first method involves studying a transient solution numerically using a very large number of time steps, which can be computationally expensive.  The second method is advantageous from the point of view that it does not require any analytical or numerical solution of the transient flow model.  However, the key limitation of the scaling argument approach is that it provides no insight into how the response time varies spatially or with the boundary conditions, nor is it obvious how the scaling argument applies to flow in heterogeneous porous media where $D$ might vary with position. Moreover, it is unclear how to choose the constant of proportionality between the response time and $L^{2}/D$.
%the first method relies on identifying a tolerance, and the choice of this tolerance is unclear and not standardised.

In this work, we use a different approach based on the concept of mean action time~\citep{mcnabb_1991,ellery_2012a,ellery_2012b}.  The benefit of working with this framework is that it avoids the need for solving the underlying parabolic partial differential equation model of transient flow, and it provides explicit information about how the response time varies with position~\citep{simpson_2013,jazaei_2014}.  In general, the mean action time approach relies on identifying a cumulative distribution function, $F(t;x)$, which varies from $F(0;x) = 0$ to $F(t;x) \to 1^{-}$ as $t \to \infty$.  Previous studies using the concept of mean action time have examined the first one or two central moments of the associated probability density function, $f(t;x) = \textrm{d}F(t;x)/\textrm{d}t$~\citep{simpson_2013,jazaei_2014}.  The first central moment is known as the \textit{mean action time} and the second central moment is known as the \textit{variance of action time}.  In this work we take a more general approach and present a new method that can be used to calculate the $k$th raw moment of $f(t;x)$ and show how to combine these moments to produce a highly accurate estimate of the response time. The formulation extends the mathematical results for diffusion in a homogeneous medium without a source/sink term, developed by \citet{carr_2017},  to a model of saturated flow through a heterogeneous porous medium with a general source/sink term to describe recharge processes~\citep{simpson_2013}. In practice, our approach can be used to arrive at a very accurate estimate of the response time using just two consecutive raw moments, $k-1$ and $k$, for a suitably large choice of $k$.  The key benefit of our approach is that it leads to more accurate estimates of the response time compared to previous estimates based on the first two central moments only~\citep{simpson_2013,jazaei_2014}.   Furthermore, the computational effort required to solve for the first few moments is far less than the computational effort required to solve for the underlying transient solution.

The new estimate of the response time is applied to two case studies.  The first case study involves flow through a homogeneous porous medium and considers both recharge and discharge processes, where the saturated thickness increases and decreases over time, respectively.  In the first case study, we use a laboratory-scale experimental data set to illustrate the strengths of our new approach~\citep{simpson_2013}.   In the second case study we examine a more practical scenario involving flow through heterogeneous porous media, where the saturated hydraulic conductivity varies spatially.  In the second case study we use numerical solutions of the governing flow equation to assess the accuracy of our estimates of the response time.  Overall, we find that the new method is both highly accurate and highly efficient. Moreover, all test cases lead to improved estimates of the response time, with practically-useful estimates, accurate to two significant digits, obtained using only the first two raw moments of $f(t;x)$.

\cbl Our approach relies upon reformulating the transient response in a groundwater flow system in terms of a cumulative distribution function, and then examining certain properties of that transient response in terms of the moments of the associated probability density function.  The goal of the analysis is to make a clear and unambiguous  distinction between transient flow conditions and effectively steady state conditions.  It is worthwhile pointing out that analyzing data and mathematical models using moment analysis and moment-based techniques is relatively common in the groundwater hydrology literature.  Methods based on the analysis of moments are used to interpret field data \citep{Asmuth2007,Besbes1984,Shapoori2015,Berendrecht2016} and to calibrate mathematical models of groundwater flow to match observed data~\citep{Bakker2007,Bakker2008}.  However, the purpose of the current study and these previous studies are very different.  Here our primary focus is not on calibrating models or interpreting field data, instead we are concerned with determining a duration of time required for a transient groundwater response to take place. \cb

This manuscript is organised in the following way. In Section \ref{sec:mathematical_formulation} we present, and analyze the governing groundwater flow equation in a one-dimensional heterogeneous aquifer with arbitrary recharge.  To analyze the problem we consider an equivalent semi-discrete formulation of the mathematical model, and we make use of the properties of the semi-discrete formulation to arrive at an approximate relationship between the response time and two consecutive moments, $k-1$ and $k$, for sufficiently large $k$.   We explain how to solve for the two consecutive moments, and provide a mathematically rigorous connection between our new theory and existing scaling approaches for a simplified problem of flow in homogeneous porous media.  Case studies to demonstrate the efficacy of our approach are presented in Section \ref{sec:results}.  Finally, in Section \ref{sec:conclusions}, we summarise the contribution of this study.
%\begin{landscape}
\begin{figure}
  \includegraphics[width=0.9\textwidth]{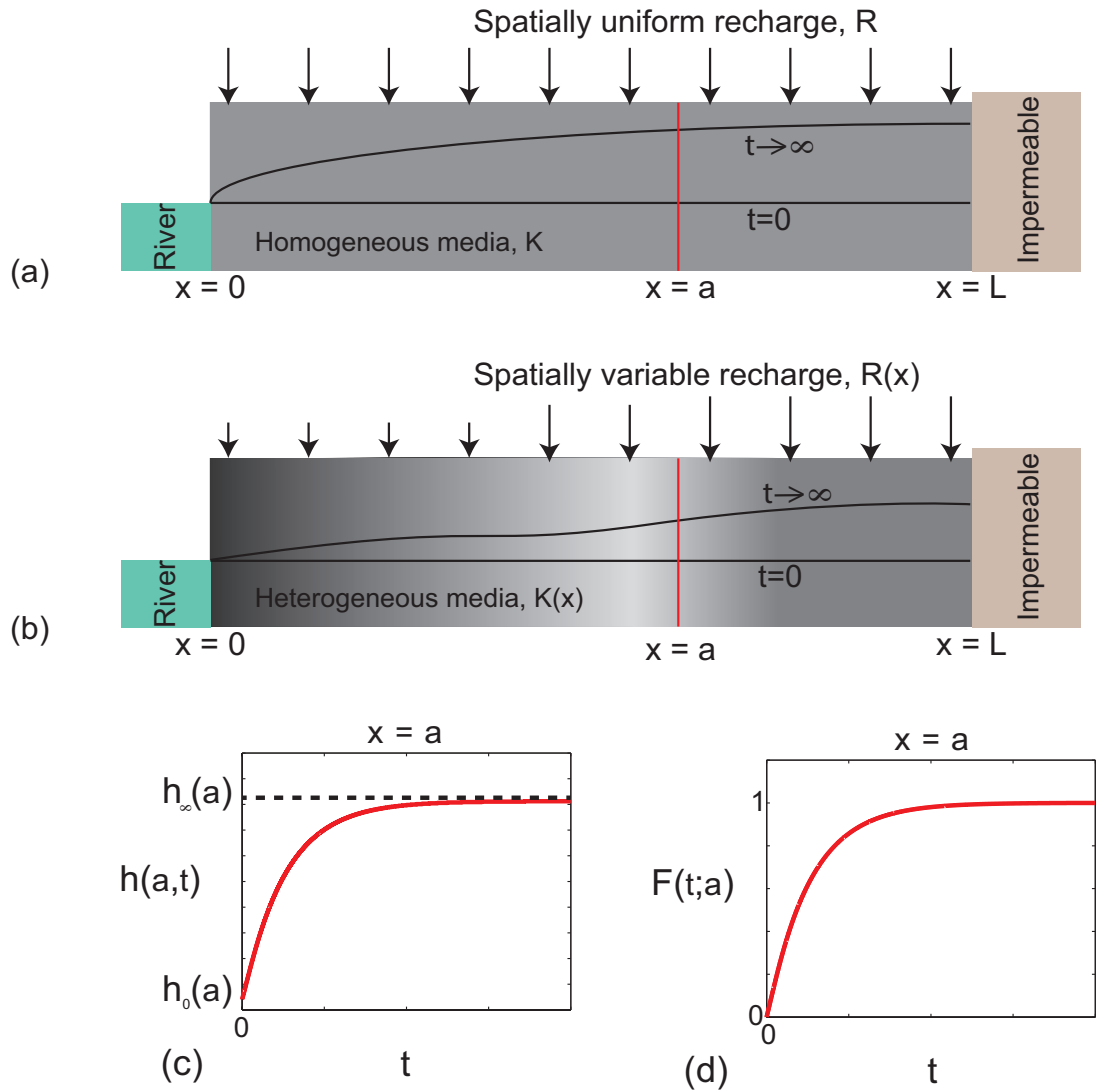}
       % Give a unique label
\caption{\selectfont{}Schematic illustrating groundwater response in: (a) a homogeneous porous medium with constant recharge; and (b) a heterogeneous porous medium with spatially variable recharge. In both schematics flow takes place in a one-dimensional aquifer, $0 < x <  L$.  At $x=0$ there is a constant saturated thickness, $h(0,t)=h_{1}$, and there is an impermeable boundary at $x=L$, giving $\partial h(L,t) / \partial x = 0$.  The initial saturated thickness is constant, $h(x,0)=h_{1}$, and a transient response takes place by the application of some recharge, $R(x)$.  In (a) flow takes place in a homogeneous porous medium with $K(x) \equiv K$ and the recharge is spatially uniform $R(x) \equiv R$.  In (b) flow takes place in a heterogeneous porous medium with a spatially varying saturated hydraulic conductivity, $K(x)$, and the recharge is spatially variable, $R(x)$.  At each location $x=a$, for $0 < a  < L$, the transient response sees $h(a,t)$ change monotonically from $h(a,0) = h_{0}(a)$ to $\lim_{t \to \infty}h(a,t) = h_{\infty}(a)$ as $t$ increases, as shown in (c).  Using the data in (c) we can construct a cumulative distribution function, $F(t;a) = 1 - (h(a,t)-h_{\infty}(a))/(h_{0}(a)-h_{\infty}(a))$, which increases from $F(0;a)=0$ and approaches $F(0,a)=1^-$ as $t \to \infty$, as illustrated in (d).   }
\label{fig:schematic}
\end{figure}
%\end{landscape}

\section{Mathematical formulation}
\label{sec:mathematical_formulation}
\subsection{Groundwater flow model}

We consider a one-dimensional Dupuit-Forchheimer model of saturated flow through a heterogeneous porous medium with a general source/sink term to describe recharge/discharge processes~\citep{Bear1972,Bear1979},
\begin{gather}
\label{eq:df_model_nonlinear}
S\frac{\partial h}{\partial t} = \frac{\partial}{\partial x}\left(K(x)h\frac{\partial h}{\partial x}\right) + R(x),\quad 0 < x < L,\enspace t > 0,\\
h(x,0) = h_{0}(x),\quad h(0,t) = h_{1},\quad \frac{\partial h}{\partial x}(L,t) = 0,
\end{gather}
where $h(x,t) > 0$ is the saturated thickness at position $x$ and time $t$, $S > 0$ is the storage coefficient, $K(x)>0$ is the spatially-varying saturated hydraulic conductivity and $R(x)$ is the spatially-varying recharge rate (Figure \ref{fig:schematic}). For many practical problems, the hydraulic gradient is small, $|\partial h/\partial x|\ll 1$, and so we can approximate Equation (\ref{eq:df_model_nonlinear}) by a linearised model~\citep{Bear1972,Bear1979}, given by
\begin{gather}
\label{eq:df_model_linear_eq}
\frac{\partial h}{\partial t} = \frac{\partial}{\partial x}\left(D(x)\frac{\partial h}{\partial x}\right) + W(x),\quad 0 < x < L,\enspace t > 0,\\
\label{eq:df_model_linear_IBCs}
h(x,0) = h_{0}(x),\quad h(0,t) = h_{1},\quad \frac{\partial h}{\partial x}(L,t) = 0,
\end{gather}
with diffusivity $D(x) = \overline{h}K(x)/S$, source term $W(x) = R(x)/S$ and average saturated thickness $\overline{h}$.  In this work, we study properties of Equations (\ref{eq:df_model_linear_eq}) and (\ref{eq:df_model_linear_IBCs}), which means that our analysis is relevant for confined flow~\citep{Bear1972,Bear1979} or unconfined flow with a sufficiently small hydraulic gradient.  If the model is used to study confined flow then the storage coefficient is the specific storage~\citep{Bear1972,Bear1979}, whereas if the model is used to study unconfined flow the storage coefficient is the specific yield~\citep{Bear1972,Bear1979}.  The initial condition is arbitrary, and the boundary conditions correspond to a constant saturated thickness at $x=0$, and a no flow boundary condition at $x=L$.  We focus on these boundary conditions because they are relevant to the laboratory data presented by \citet{simpson_2013}, however, our approach can be adapted to deal with other boundary conditions, and we will explain these details later (Section \ref{sec:general_boundary_conditions}).  Furthermore, all our theoretical results are demonstrated here using a one-dimensional model so that we can present the theory as simply as possible.  However, all developments also apply to two- and three-dimensional flow, and we will make a comment on how these generalisations can be implemented later \citep{Simpson2018}.

\subsection{Definition of response time}
Our aim is to calculate a \textit{response time}, or a measure of the time required for the transient solution of Equations (\ref{eq:df_model_linear_eq}) and (\ref{eq:df_model_linear_IBCs}), $h(x,t)$, to effectively reach the corresponding steady state solution, which can be written as $\displaystyle{h_{\infty}(x) = \lim_{t \to \infty} h(x,t)}$, and which satisfies the elliptic problem
%\begin{subequations}
%\label{eq:df_model_linear_ss}
\begin{gather}
\label{eq:df_model_linear_eq_ss}
0 = \frac{\textrm{d}}{\textrm{d} x}\left(D(x)\frac{\textrm{d} h_{\infty}}{\textrm{d} x}\right) + W(x),\quad 0 < x < L,\\
\label{eq:df_model_linear_IBCs_ss}
h_{\infty}(0) = h_{1},\quad \frac{\textrm{d} h_{\infty}}{\textrm{d} x}(L) = 0,
\end{gather}
%\end{subequations}
For each position $x$, we take the response time to be the time,  $t = t_{r}$, satisfying
\begin{gather}
\label{eq:finite_response_time}
\frac{h(x,t_{r}) - h_{\infty}(x)}{h_{0}(x) - h_{\infty}(x)} = \delta,
\end{gather}
where $\delta\ll 1$ is some sufficiently small, prescribed tolerance specifying the proportion of the transition from $h_{0}(x)$ to $h_{\infty}(x)$ remaining at $t = t_{r}$. In general, we expect that $t_{r}$ will vary with position, giving $t_{r}(x)$.  In a practical sense, we can simply define the global response time to be the maximum value of $t_{r}(x)$ \citep{carr_2017}.  Furthermore, the choice of $\delta$ can be guided by the accuracy with which measurements of the hydraulic head can be made in a practical setting.

The aim of this work is to calculate highly accurate estimates of the response time, $t_{r}$, without computing the transient solution of Equations (\ref{eq:df_model_linear_eq}) and (\ref{eq:df_model_linear_IBCs}), $h(x,t)$. To make progress we examine the following quantity
\begin{gather}
\label{eq:CDF}
F(t;x) = 1 - \frac{h(x,t) - h_{\infty}(x)}{h_{0}(x) - h_{\infty}(x)},
\end{gather}
which, for many practical flow problems, varies monotonically from $F(0;x)=0$ to $F(t;x) \to 1^{-}$ in the long time limit as $t \to \infty$. Combining Equations (\ref{eq:finite_response_time}) and (\ref{eq:CDF}), the response time $t_{r}$ equivalently satisfies
\begin{gather}
\label{eq:CDF_tolerance}
F(t_{r}; x) = 1 - \delta.
\end{gather}
We can now reformulate the problem into a common one from probability theory, by observing that $F(t; x)$ can be interpreted as a cumulative distribution function of time $t$, at each location $x$.  This will always be the case provided that $h(x,t)$ varies monotonically from $h_{0}(x)$ to $h_{\infty}(x)$.  In addition to Equation (\ref{eq:CDF}), our analysis will also make use of the corresponding probability density function, $f(t; x)$, given by
\begin{gather}
\label{eq:pdf}
f(t;x) = \frac{\textrm{d} F(t;x)}{\textrm{d} t} = \frac{1}{h_{\infty}(x)-h_{0}(x)}\frac{\partial}{\partial t}\left[h(x,t) - h_{\infty}(x)\right].
\end{gather}

\subsection{Studying the large time behaviour}
Since we are only interested in small values of $\delta$, only the large time behaviour of $F(t; x)$ is important when estimating $t_{r}$. To examine this behaviour, according to Equation (\ref{eq:CDF}), we must study the behaviour of the transient solution, $h(x,t)$, for large $t$. This can be achieved by first carrying out a spatial-discretization of Equations (\ref{eq:df_model_linear_eq}) and (\ref{eq:df_model_linear_IBCs}), say on a grid with nodes at $x=x_{i}$ ($0 = x_{0} < x_{1} < \hdots < x_{n} = L$), using a standard finite difference, finite volume or finite element method~\citep{Wang1983,Zheng2002}.  After carrying out the spatial discretization we can study the time-dependent behaviour of the resulting semi-discrete system, which can be expressed as~\citep{barry_1996,simpson_2007}
\begin{gather}
\label{eq:ode_system}
\frac{\textrm{d}\mathbf{h}}{\textrm{d}t} = -\mathbf{A}\mathbf{h} + \mathbf{b},\quad \mathbf{h}(0) = \mathbf{h}_{0},
\end{gather}
where $\mathbf{h} = (h_{1}(t),\hdots,h_{n}(t))^{T}$ is the spatially discretized transient solution, with $h_{i}(t)$ denoting the numerical approximation of $h(x,t)$ at $x = x_{i}$ and time $t$, $\mathbf{h}_{0} = (h_{0}(x_{1}),\hdots,h_{0}(x_{n}))^{T}$ is the spatially discretized initial condition, $\mathbf{A}$ is a \cbl positive definite \cb matrix representing the spatial discretization of the linear operator $\mathcal{L}$, defined by $\mathcal{L}h = \frac{\partial}{\partial x}\left(-D(x)\frac{\partial h}{\partial x}\right)$, and $\mathbf{b}$ is a vector that incorporates both the boundary conditions and the spatial discretization of the source term, $W(x)$. We remark that a semi-discrete equation is not included at the left boundary ($x=x_{0}$) in Equation (\ref{eq:ode_system}) since $h(0,t) = h_{1}$ for all time $t$.  A key strength of our analysis is that we can deal with a very general mathematical model which describes flow in a heterogeneous porous medium where both the aquifer diffusivity, $D(x)$, and the recharge rate, $R(x)$, can take on arbitrary functional forms.  Despite the fact that exact solutions of the partial differential equation description of such general problems are rarely available, except for quite particular forms of $D(x)$ and $R(x)$~\citep{Zoppou1997,Zoppou1999}, we can still make analytical progress with our approach.  The key to making progress is that we study the time dependent solution of the semi-discrete model in the long-time limit, as $t \to \infty$.

The exact solution of Equation (\ref{eq:ode_system}) is
\begin{gather}
\label{eq:numerical_pde_solution}
\mathbf{h}(t) = \mathbf{h}_{\infty} + e^{-t\mathbf{A}}\left(\mathbf{h}_{0} - \mathbf{h}_{\infty}\right),
\end{gather}
where $\mathbf{h}_{\infty} = -\mathbf{A}^{-1}\mathbf{b}$ is the spatially discretized steady state solution. \cbl Using the diagonalization of $\mathbf{A}$~\citep{Wilkinson1965}\cb, we can express the matrix exponential as
\begin{equation}
\label{eq:matrix_exponential}
e^{-t\mathbf{A}} = \sum_{j=1}^{n}e^{-t\lambda_{j}}\mathbf{x}_{j}\mathbf{y}_{j}^{T},
\end{equation}
where $\lambda_{1},\hdots,\lambda_{n}$ are the positive eigenvalues of $\mathbf{A}$ arranged in order of increasing magnitude, $\mathbf{x}_{1},\hdots,\mathbf{x}_{n}$ are the corresponding eigenvectors and $\mathbf{y}_{j}^{T}$ is the $j$th row of the inverse of the matrix $\left[\mathbf{x}_{1},\hdots,\mathbf{x}_{n}\right]$. Substituting Equation (\ref{eq:matrix_exponential}) into Equation (\ref{eq:numerical_pde_solution}) yields
\begin{equation}
\label{eq:numerical_pde_solution_analytic}
\mathbf{h}(t) = \mathbf{h}_{\infty} + \sum_{j=1}^{n}\mathbf{v}_{j}e^{-t\lambda_{j}},
\end{equation}
where $\mathbf{v}_{j} = \mathbf{x}_{j}\mathbf{y}_{j}^{T}\left(\mathbf{h}_{0} - \mathbf{h}_{\infty}\right)$.
%Substituting (\ref{eq:matrix_exponential}) into (\ref{eq:numerical_pde_solution}) we obtain the following large time behaviour of (\ref{eq:numerical_pde_solution}):
%\begin{equation}
%\label{eq:numerical_pde_solution_asymptotic}
%\mathbf{h}(t) \simeq \mathbf{h}_{\infty} + \mathbf{x}_{1}\mathbf{y}_{1}^{T}\left(\mathbf{h}_{0} - \mathbf{h}_{\infty}\right)e^{-t\lambda_{1}}
%\end{equation}
Therefore, at each position $x$, the solution, Equation (\ref{eq:numerical_pde_solution_analytic}), can be written as
%\begin{equation}
%h_{i}(t) \simeq h_{\infty}(x_{i}) + \alpha_{i}e^{-t\beta_{i}}
%\end{equation}
%for large time $t$, where $\alpha_{i}$ and $\beta_{i}$ are time-independent constants that depend on the position $x_{i}$.
%\begin{equation}
%h(x,t) \simeq h_{\infty}(x) + \alpha(x)e^{-t\beta(x)}
%\end{equation}
\begin{equation}
h(x,t) = h_{\infty}(x) + \sum_{j=1}^{n}v_{j}(x) e^{-t\lambda_{j}}.
\label{eq:functional_form}
\end{equation}
Substituting Equation (\ref{eq:functional_form}) into Equation (\ref{eq:CDF}) leads to a specific functional form for the cumulative distribution function
\begin{gather}
\label{eq:CDF_functional_form}
F(t;x) = 1 - \sum_{j=1}^{n}\alpha_{j}(x)e^{-t\lambda_{j}},
\end{gather}
where $\alpha_{j}(x) = v_{j}(x)/(h_{0}(x)-h_{\infty}(x))$. Since $\lambda_{1}$ is the smallest eigenvalue, $e^{-t\lambda_{1}} \gg e^{-t\lambda_{j}}$ for large $t$ and for all $j = 2,\hdots,n$, and hence:
\begin{equation}
\label{eq:CDF_larget}
F(t; x) \sim 1 - \alpha_{1}(x)e^{-t\lambda_{1}}, \quad  \text{for large $t$}.
\end{equation}

\subsection{Simple formula for the response time}% for flow in heterogeneous porous media}
Substituting Equation (\ref{eq:CDF_larget}) into Equation (\ref{eq:CDF_tolerance}) and solving for $t_{r}$ yields the following approximation for the response time
\begin{equation}
\label{eq:tdash_asymptotic}
t_{r} \sim \frac{1}{\lambda_{1}}\log_{\textrm{e}}\left(\frac{\alpha_{1}(x)}{\delta}\right),\quad\text{for small $\delta$}.
\end{equation}
This asymptotic expression for the response time depends on two unknown parameters, $\lambda_{1}$ and $\alpha_{1}(x)$, however, as shown previously by \citet{carr_2017} for the homogeneous diffusion equation without any source term or spatial variations in the coefficients, we can compute these parameters using the raw moments of the underlying probability distribution. The $k$th raw moment at position $x$ is
\begin{equation}
\label{eq:raw_moment}
M_{k}(x) = \int_{0}^{\infty}t^{k}f(t;x)\,\textrm{d}t,\quad k = 0,1,2,\hdots,
\end{equation}
where $f(t;x)$ is the probability density function, obtained by differentiating Equation (\ref{eq:CDF_functional_form}) with respect to time $t$
\begin{gather}
f(t;x) = \sum_{j=1}^{n}\alpha_{j}(x)\lambda_{j}e^{-t\lambda_{j}}.
\end{gather}
Substituting this expression for $f(t; x)$ into Equation (\ref{eq:raw_moment}) leads to
\begin{equation*}
M_{k}(x) = \sum_{j=1}^{n}\lambda_{j}\alpha_{j}(x)\int_{0}^{\infty}t^{k}e^{-t\lambda_{j}}\,\textrm{d}t.
\end{equation*}
Using integration by parts repeatedly, and noting that $\displaystyle{\lim_{t\rightarrow\infty} t^{m}e^{-t\lambda_{j}}} = 0$ for $m = 1,\hdots,k$, one can show that
\begin{equation*}
\int_{0}^{\infty}t^{k}e^{-t\lambda_{j}}\,dt = \frac{k!}{\lambda_{j}^{k+1}},
\end{equation*}
and hence
\begin{equation*}
M_{k}(x) = k!\sum_{j=1}^{n}\frac{\alpha_{j}(x)}{\lambda_{j}^{k}}.
\end{equation*}
Since $\lambda_{1}$ is the smallest eigenvalue, $\lambda_{1}^{k}\ll \lambda_{j}^{k}$ for large $k$ and for all $j = 2,\hdots,n$ and hence we have the asymptotic result:
\begin{equation}
M_{k}(x) \sim \frac{\alpha_{1}(x)k!}{\cbl\lambda\cb_{1}^{k}},\quad\text{for large $k$},
\end{equation}
which relates the two unknown parameters in Equation (\ref{eq:tdash_asymptotic}), $\lambda_{1}$ and $\alpha_{1}(x)$, to the $k$th moment. For suitably large $k$, using two consecutive moments, we have
\begin{gather}
M_{k-1}(x) \approx \frac{\alpha_{1}(x)(k-1)!}{\cbl\lambda\cb_{1}^{k-1}},\quad
M_{k}(x) \approx \frac{\alpha_{1}(x)k!}{\cbl\lambda\cb_{1}^{k}},
\end{gather}
which can be solved to give
\begin{equation}
\label{eq:alpha_beta}
\alpha_{1}(x) \approx \frac{M_{k}(x)}{k!}\left(\frac{kM_{k-1}(x)}{M_{k}(x)}\right)^{k},\quad
\lambda_{1}(x) \approx \frac{kM_{k-1}(x)}{M_{k}(x)}.
\end{equation}
Inserting the approximations given by Equation (\ref{eq:alpha_beta}) into Equation (\ref{eq:tdash_asymptotic}) yields the following estimate of the response time
\begin{equation}
\label{eq:response_time}
\mathrm{RT}(x; k,\delta) = \frac{M_{k}(x)}{kM_{k-1}(x)}\log_{\textrm{e}}\left[\frac{M_{k}(x)}{k! \,\delta}\left(\frac{kM_{k-1}(x)}{M_{k}(x)}\right)^{k}\right],
\end{equation}
where we introduce the notation $\mathrm{RT}(x; k,\delta)$ for the response time at position $x$. Note that $\mathrm{RT}(x; k,\delta)$ also depends on the choice of $k$ and $\delta$, which we treat as parameters. As we will demonstrate later, $\mathrm{RT}(x; k,\delta)$ provides a highly accurate estimate of the true response time at position $x$, $t_{r}(x)$, provided $k$ is sufficiently large. While the moments depend on the initial and steady state solutions, $h_{0}(x)$ and $h_{\infty}(x)$, remarkably $\mathrm{RT}(x; k,\delta)$ does not depend on the transient solution, $h(x,t)$.  This results in large computational savings as it means that we can compute the response time without solving the underlying transient flow problem formulated in Equations (\ref{eq:df_model_linear_eq}) and (\ref{eq:df_model_linear_IBCs}).

Later, we will compare $\mathrm{RT}(x; k,\delta)$ to two other measures of the time required to reach steady state, $\mathrm{MAT}(x)$ and $(\mathrm{MAT} + \sqrt{\mathrm{VAT}})(x)$, corresponding to the mean of the probability density function, and the mean plus one standard deviation of the probability density function $f(t; x)$~\citep{ellery_2013}.  These quantities involve the computation of the first two central moments, that are known as the mean action time (MAT) and the variance of action time (VAT), respectively.  These two estimates can be expressed in terms of the first two raw moments, $M_{1}(x)$ and $M_{2}(x)$, as
\begin{gather}
\label{eq:MAT}
\mathrm{MAT}(x) = M_{1}(x),\\
\label{eq:MATVAT}
(\mathrm{MAT} + \sqrt{\mathrm{VAT}})(x) = M_{1}(x) + \sqrt{M_{2}(x) - M_{1}^{2}(x)}.
\end{gather}
Later when we consider applying our new theory to certain case studies we will compare the accuracy of $\mathrm{RT}(x; k,\delta)$, $\mathrm{MAT}(x)$ and $(\mathrm{MAT} + \sqrt{\mathrm{VAT}})(x)$.

\subsection{Computation of the raw moments}
\label{sec:moments}
To apply the new method, all that is required to compute $\mathrm{RT}(x; k,\delta)$ is to supply the raw moments in Equation (\ref{eq:response_time}). To achieve this we present a new algorithm that extends the approach of \citet{carr_2017} to a spatially-varying diffusivity and non-zero source term. Using integration by parts, the $k$th raw moment, Equation (\ref{eq:raw_moment}), can be expressed as \citep{carr_2017,simpson_2013}
\begin{equation}
\label{eq:raw_moment_alternative}
M_{k}(x) = \frac{k}{g(x)}\int_{0}^{\infty} t^{k-1}\left[h_{\infty}(x)-h(x,t)\right]\,\textrm{d}t,
\end{equation}
where we define $g(x) = h_{\infty}(x) - h_{0}(x)$ to keep the notation succinct. Next, define $\overline{M}_{k}(x) = M_{k}(x)g(x)$, and hence from Equation (\ref{eq:raw_moment_alternative}) we have
\begin{equation}
\label{eq:Mbar}
\overline{M}_{k}(x) = k\int_{0}^{\infty} t^{k-1}\left[h_{\infty}(x)-h(x,t)\right]\,\textrm{d}t.
\end{equation}
Differentiating both sides of Equation (\ref{eq:Mbar}) with respect to $x$, multiplying the result by $D(x)$ and then differentiating again with respect to $x$, yields
\begin{equation*}
\frac{\textrm{d}}{\textrm{d}x}\left(D(x)\frac{\textrm{d}\overline{M}_{k}}{\textrm{d}x}\right) = k\int_{0}^{\infty} t^{k-1}\left[\frac{\textrm{d}}{\textrm{d}x}\left(D(x)\frac{\textrm{d}h_{\infty}}{\textrm{d}x}\right)-\frac{\partial}{\partial x}\left(D(x)\frac{\partial h}{\partial x}\right)\right]\,\textrm{d}t.
\end{equation*}
The source term $W(x)$ can be added and subtracted in the integrand to give
\begin{equation}
\label{eq:ode_formulation1}
\frac{\textrm{d}}{\textrm{d}x}\left(D(x)\frac{\textrm{d}\overline{M}_{k}}{\textrm{d}x}\right) = k\int_{0}^{\infty} t^{k-1}\left[\frac{\textrm{d}}{\textrm{d}x}\left(D(x)\frac{\textrm{d}h_{\infty}}{\textrm{d}x}\right)+W(x)-\frac{\partial}{\partial x}\left(D(x)\frac{\partial h}{\partial x}\right)-W(x)\right]\,\textrm{d}t.
\end{equation}
Now, noting the form of Equation (\ref{eq:df_model_linear_eq}), Equation (\ref{eq:ode_formulation1}) simplifies further to:
\begin{equation}
\label{eq:ode_formulation2}
\frac{\textrm{d}}{\textrm{d}x}\left(D(x)\frac{\textrm{d}\overline{M}_{k}}{\textrm{d}x}\right) = k\int_{0}^{\infty} t^{k-1}\frac{\partial}{\partial t}\left[h_{\infty}(x)-h(x,t)\right]\,\textrm{d}t.
\end{equation}
From Equations (\ref{eq:pdf}) and (\ref{eq:raw_moment}), the right-hand side of Equation (\ref{eq:ode_formulation2}) can be expressed in terms of the $(k-1)$th raw moment. In summary, the $k$th moment satisfies the following boundary value problem
%\begin{subequations}
%\label{eq:moments_bvp}
\begin{gather}
\label{eq:moments_bvp_ode}
\frac{\textrm{d}}{\textrm{d}x}\left(D(x)\frac{\textrm{d}\overline{M}_{k}}{\textrm{d}x}\right) = -k\overline{M}_{k-1}(x),\quad 0 < x < L,\\
\label{eq:moments_bvp_bcs}
\overline{M}_{k}(0) = 0,\quad \frac{\textrm{d}\overline{M}_{k}}{\textrm{d}x}(L) = 0.
\end{gather}
%\end{subequations}
The boundary conditions, Equation (\ref{eq:moments_bvp_bcs}), are derived using the boundary conditions given by Equation (\ref{eq:df_model_linear_IBCs}) together with appropriate expressions for $\overline{M}_{k}(x)$ and $\textrm{d}\overline{M}_{k}/\textrm{d}x$. For example, combining $h_{\infty}(0) = h_{1}$ and $h(0,t) = h_{1}$ with Equation (\ref{eq:Mbar}) gives the stated boundary condition at $x = 0$.

The boundary value problem, Equations (\ref{eq:moments_bvp_ode}) and (\ref{eq:moments_bvp_bcs}), defines a recursive relation between the functions $\overline{M}_{0}(x)$, $\overline{M}_{1}(x)$ and so on. Starting with $k=1$, the right-hand side of Equation (\ref{eq:moments_bvp_ode}) is known, with $\overline{M}_{0}(x) = M_{0}(x)g(x) = g(x)$, since the zeroth order moment $M_{0}(x) = 1$, so Equations (\ref{eq:moments_bvp_ode}) and (\ref{eq:moments_bvp_bcs}) can be solved for $\overline{M}_{1}(x)$. With $\overline{M}_{1}(x)$ now known, the boundary value problem is solved for $\overline{M}_{2}(x)$. This process is repeated until a desired order. At each step, the raw moment, which ultimately appears in the response time formula in Equation (\ref{eq:response_time}), is simply given by $M_{k}(x) = \overline{M}_{k}(x)/g(x)$.

\cbl
\subsection{Extension to other types of boundary conditions}
\label{sec:general_boundary_conditions}
To present the new theory as clearly as possible, all results are developed for the boundary conditions in Equation (\ref{eq:df_model_linear_IBCs}) because these boundary conditions are relevant to the laboratory data set presented by~\cite{simpson_2013}.  However, more general boundary conditions are easily accommodated.   For example, if we consider the general boundary conditions
\begin{gather}
\label{eq:general_bc_h}
a_{0}h(0,t) - b_{0}\frac{\partial h}{\partial x}(0,t) = c_{0},\quad a_{L}h(L,t) + b_{L}\frac{\partial h}{\partial x}(L,t) = c_{L},
\end{gather}
where $a_{0}$, $b_{0}$, $c_{0}$, $a_{L}$, $b_{L}$, $c_{L}$ are specified constants, the formula for the response time, Equation (\ref{eq:response_time}), still holds. However, the boundary conditions appearing in Equation (\ref{eq:moments_bvp_bcs}) become
\begin{gather}
\label{eq:moments_bvp_bcs_general}
a_{0}\overline{M}_{k}(0) - b_{0}\frac{\textrm{d}\overline{M}_{k}}{\textrm{d}x}(0) = 0,\quad a_{L}\overline{M}_{k}(L) + b_{L}\frac{\textrm{d}\overline{M}_{k}}{\textrm{d}x}(L) = 0.
\end{gather}
These general boundary condition are obtained using
\begin{gather*}
a_{0}\overline{M}_{k}(x) - b_{0}\frac{\textrm{d}\overline{M}_{k}}{\textrm{d}x}(x) = k\int_{0}^{\infty} t^{k-1}\left[a_{0}h_{\infty}(x)-b_{0}h_{\infty}'(x)-\left(a_{0}h(x,t) - b_{0}\frac{\partial h}{\partial x}(x,t)\right)\right]\,\textrm{d}t,\\
a_{L}\overline{M}_{k}(x) + b_{L}\frac{\textrm{d}\overline{M}_{k}}{\textrm{d}x}(x) = k\int_{0}^{\infty} t^{k-1}\left[a_{L}h_{\infty}(x)+b_{L}h_{\infty}'(x)-\left(a_{L}h(x,t) + b_{L}\frac{\partial h}{\partial x}(x,t)\right)\right]\,\textrm{d}t,
\end{gather*}
which are formed by taking appropriate linear combinations of the integral expression for $\overline{M}_{k}(x)$, Equation (\ref{eq:Mbar}), and its derivative with respect to $x$. Substituting $x = 0$ and $x = L$ into these expressions, respectively, and noting the form of the boundary conditions satisfied by $h(x,t)$, Equation (\ref{eq:general_bc_h}), and the corresponding boundary conditions satisfied by $h_{\infty}(x)$, produces the stated results, Equation (\ref{eq:moments_bvp_bcs_general}).
\cb

\subsection{Computational cost}
Calculating the  response time, Equation (\ref{eq:response_time}), for $k = m$, requires the solution of  $m+1$ boundary value problems.  These boundary value problems include solving Equations (\ref{eq:df_model_linear_eq_ss}) and (\ref{eq:df_model_linear_IBCs_ss}) for $h_{\infty}(x)$, and solving Equations (\ref{eq:moments_bvp_ode}) and (\ref{eq:moments_bvp_bcs}) $m$ times for  the moments $M_{k}(x)$, $k = 1,\hdots,m$.  In this work we solve both boundary value problems using a vertex-centered finite volume method on a uniform grid utilizing a second-order central difference approximation to the first-order spatial derivatives \citep{eymard_2000}.  For each boundary value problem, the spatial discretization yields a system of linear equations, of size $n \times n$, where $n$ is the number of unknown values as defined in Equation (\ref{eq:ode_system}).  These $m+1$ linear systems are then solved to obtain the discrete numerical approximations to $h_{\infty}(x)$ and $M_{k}(x)$ for $k = 1,\hdots,m$. In summary, the computational cost of computing the response time using Equation (\ref{eq:response_time}) for $k = m$ is the solution of $m+1$ linear systems of size $n \times n$.

In comparison, calculating the response time via Equation (\ref{eq:finite_response_time}) by computing the transient solution of Equations (\ref{eq:df_model_linear_eq}) and (\ref{eq:df_model_linear_IBCs}) is always significantly more computationally expensive.  Using the same finite volume discretization and a standard backward/implicit Euler temporal scheme, as is often employed in MODFLOW~\citep{harbaugh_2005}, to solve Equations (\ref{eq:df_model_linear_eq}) and (\ref{eq:df_model_linear_IBCs}) requires the solution of a linear system of size $n \times n$ at each time step.   If a time dependent solution of Equations (\ref{eq:df_model_linear_eq}) and (\ref{eq:df_model_linear_IBCs}) is used to study the response time, then the numerical solution for $h(x,t)$ must be obtained for sufficiently large $t$ to examine the response time.  Furthermore, to control truncation error, the value of the time step must be sufficiently small.  These two requirements mean that a very large number of time steps are required to estimate the response time by solving Equations (\ref{eq:df_model_linear_eq}) and (\ref{eq:df_model_linear_IBCs}).

\cbl
As we demonstrate in Section \ref{sec:results}, very accurate estimates of the response time, $t_{r}(x)$, can be computed via Equation (\ref{eq:response_time}) with $k = m$, for $m = \text{3, 4 or 5}$, giving $m+1 = \text{4, 5 or 6}$.   This means that the new method can be used by solving just four to six linear systems, and this number dwarfs the number of time steps (and hence linear systems) required in a typical transient simulation by at least an order of magnitude, and possibly several orders of magnitude~\citep{Simpson2018}. Therefore, the new response time formula $\mathrm{RT}(x;k,\delta)$ offers significant computational savings. Although our approach does not supply the transient solution, $h(x,t)$, it provides an efficient and accurate alternative to solving the groundwater flow model, Equations (\ref{eq:df_model_linear_eq}) and (\ref{eq:df_model_linear_IBCs}), for calculating the response time. \cb

\subsection{A rigorous connection to scaling results for flow in homogeneous porous media}
\label{sec:simple_formula_homogeneous}
As stated in the Introduction, one common approach to estimate the response time for flow in homogeneous porous media is to claim that the response time is proportional to $L^2/D$, where $L$ is a relevant length scale and $D$ is the aquifer diffusivity.  While this kind of calculation is widespread in the literature~\citep{Bredehoeft_2009,Gelhar1974,haitjema2006,Manga1999}, it is difficult to use this approach in practice because this kind of argument provides no information about the appropriate constant of proportionality.  In this section we will apply our new theory to the simplified problem of flow in a homogeneous porous medium where we consider Equations (\ref{eq:df_model_linear_eq}) and (\ref{eq:df_model_linear_IBCs}) with constant values for the initial condition, saturated hydraulic conductivity and recharge rate: $h_{0}(x) = h_{1}$, $K(x)\equiv K$ and $R(x)\equiv R \neq 0$.  Together, these simplifications give $D(x)\equiv D = \overline{h}K/S$ and $W(x)\equiv W = R/S$, leading to
%\begin{subequations}
%\label{eq:df_model_linear_homogeneous}
\begin{gather}
\label{eq:df_model_linear_eq_homogeneous}
\frac{\partial h}{\partial t} = D \frac{\partial^2 h}{\partial x^2} + W,\quad 0 < x < L,\enspace t > 0,\\
\label{eq:df_model_linear_IBCs_homogeneous}
h(x,0) = h_{1},\quad h(0,t) = h_{1},\quad \frac{\textrm{d} h}{\textrm{d} x}(L,t) = 0.
\end{gather}
%\end{subequations}
For this simplified problem, we aim to derive a very simple formula for the global response time, namely $\mathrm{RT}(x; k,\delta)$, evaluated at $x = L$ where the maximum response time occurs. The key aim of this section is to examine how this estimate relates to the traditional approach of claiming that the response time is proportional to $L^2/D$.

The steady state solution of Equations (\ref{eq:df_model_linear_eq_homogeneous}) and (\ref{eq:df_model_linear_IBCs_homogeneous}) is given by
\begin{gather}
\label{eq:steady_homogeneous}
h_{\infty}(x) = \frac{Wx}{2D}(2L-x) + h_{1},
\end{gather}
and hence
\begin{gather}
\overline{M}_{0}(x) = g(x) = \frac{Wx}{2D}(2L-x).
\end{gather}
With this expression for $\overline{M}_{0}(x)$, each of the functions $\overline{M}_{k}(x)$, $k = 1,2,\hdots,m$, can be obtained recursively, using the strategy discussed at the end of Section \ref{sec:moments} and by solving the homogeneous analogue of Equation (\ref{eq:moments_bvp_ode}) by direct integration. Carrying out this process, the first few solutions evaluated at $x = L$, are given by:
\begin{align}
\label{eq:first_few_Mbars}
\hspace{-0.1cm}\overline{M}_{0}(L) = \frac{1}{2}\frac{WL^{2}}{D},\quad
\overline{M}_{1}(L) = \frac{5}{24}\frac{WL^{4}}{D^{2}},\quad
\overline{M}_{2}(L) = \frac{61}{360}\frac{WL^{6}}{D^{3}},\quad
\overline{M}_{3}(L) = \frac{1385}{6720}\frac{WL^{8}}{D^{4}},
\end{align}
suggesting that there is some kind of pattern. Using the Online Encyclopedia of Integer Sequences \citep{oeis_2017}, the numbers $1,5,61,1385$ appearing in the numerators in Equation (\ref{eq:first_few_Mbars}) are the second to fifth Euler numbers (\href{http://oeis.org/A000364}{http://oeis.org/A000364}) while the numbers $2,24,360,6720$ are the first four numbers in the integer sequence, $(2k+2)! / k!$, for $k = 0,1,2,3$ (\href{http://oeis.org/A126804}{http://oeis.org/A126804}). Therefore, we propose that
\begin{gather}
\label{eq:Mbar_general_form}
\overline{M}_{k}(L) = \frac{k!}{(2k+2)!}E_{k+2}\frac{WL^{2k+2}}{D^{k+1}},\quad\text{$k = 0,1,2,\hdots,m$},
\end{gather}
where $E_{i}$ is the $i$th positive Euler number. We have strong reason to believe this proposition as it is easily verified symbolically \citep{maple_2016} for arbitrarily large $k$.

Dividing Equation (\ref{eq:Mbar_general_form}) by $g(L) = WL^{2}/2D$ gives the $k$th moment at $x = L$:
\begin{gather}
\label{eq:moment_general_form}
M_{k}(L) = \frac{k!}{2(2k+2)!}E_{k+2}\frac{L^{2k}}{D^{k}}.
\end{gather}
Note that $W$ cancels from the numerator and denominator when arriving at Equation (\ref{eq:moment_general_form}).  This means that the moments, and hence the response time given by Equation (\ref{eq:response_time}), are independent of $W$ and therefore also independent of the recharge rate $R$. Using the asymptotic result \citep{borwein_1989}:
\begin{gather*}
E_{k+2} \sim \frac{4^{k+2}(2k+2)!}{\pi^{2k+3}},\quad\text{for large $k$},
\end{gather*}
and noting the form of Equation (\ref{eq:moment_general_form}) yields
\begin{gather}
\label{eq:moment_homogeneous_asymptotic}
M_{k}(L) \sim \frac{k!}{2}\frac{4^{k+2}}{\pi^{2k+3}}\frac{L^{2k}}{D^{k}},\quad\text{for large $k$}.
\end{gather}
Finally, substituting Equation (\ref{eq:moment_homogeneous_asymptotic}) and the equivalent form for $M_{k-1}(L)$, into the expression for $\mathrm{RT}(L; k,\delta)$, obtained by evaluating Equation (\ref{eq:response_time}) at $x = L$, produces the remarkably simple result, $\mathrm{RT}(L; k, \delta) \sim \mathrm{RT}(\delta)$ for large $k$, where
\begin{gather}
\label{eq:RTdelta_D}
\mathrm{RT}(\delta) = \frac{4}{\pi^{2}}\frac{L^{2}}{D}\log_{\mathrm{e}}\left(\frac{32}{\pi^{3}\delta}\right),
\end{gather}
and $D = \overline{h}K/S$. Note that $\mathrm{RT}(\delta)$ provides the global response time, the amount of time required for the entire transient response to effectively reach steady state. However, Equation (\ref{eq:RTdelta_D}) is only applicable to the homogeneous problem for this particular set of boundary conditions and initial conditions.

\begin{table}
\centering
\def\arraystretch{0.7}
\begin{tabular}{|cc|}
\hline
$\delta$ & $\mathrm{RT}(\delta)$\\
\hline
%\multicolumn{2}{|c|}{{}} & \multicolumn{2}{c|}{}\\[-0.4cm]
%$10^{-1}$ & $\dfrac{0.9459877999L^2}{D}$ & $10^{-4}$ & $\dfrac{3.7455955646L^2}{D}$\\[0.2cm]
%$10^{-2}$ & $\dfrac{1.8791903881L^2}{D}$ & $10^{-5}$ & $\dfrac{4.6787981528L^2}{D}$\\[0.2cm]
%$10^{-3}$ & $\dfrac{2.8123929763L^2}{D}$ & $10^{-6}$ & $\dfrac{5.6120007411L^2}{D}$\\[0.2cm]
$10^{-1}$ & $0.9460\,L^2/D$\\[0.1cm]
$10^{-2}$ & $1.8792\,L^2/D$\\[0.1cm]
$10^{-3}$ & $2.8124\,L^2/D$\\[0.1cm]
$10^{-4}$ & $3.7456\,L^2/D$\\[0.1cm]
$10^{-5}$ & $4.6788\,L^2/D$\\[0.1cm]
$10^{-6}$ & $5.6120\,L^2/D$\\
\hline
\end{tabular}
\caption{\selectfont{}Simple formulae for calculating the global response time for the homogeneous problem, Equations (\ref{eq:df_model_linear_eq_homogeneous}) and (\ref{eq:df_model_linear_IBCs_homogeneous}), with the constants of proportionality rounded to four decimal places.}% with constant values of $h_{0}(x) = h_{1}$, $K(x)\equiv K$ and $R(x)\equiv R$, $D(x)\equiv D = \overline{h}K/S_{y}$ and $W(x)\equiv W = R/S_{y}$.}
\label{tab:simple_formulae}
\end{table}
We note that the expression for $\mathrm{RT}(\delta)$ takes the form of a constant, depending on the specified tolerance $\delta$, multiplied by the time scale $L^{2}/D$. Therefore, in summary, we find that our approach provides a rigorous mathematical connection with the often stated claim that the response time is proportional to $L^2/D$~\citep{Bredehoeft_2009,Gelhar1974,haitjema2006,Manga1999}. Some very simple formulae for specific values of the tolerance $\delta$ are listed in Table \ref{tab:simple_formulae}. These results show that our approach allows us to define the constant of proportionality and we find that it varies with our choice of $\delta$, in a way that makes intuitive sense.  Namely, that the constant of proportionality increases as $\delta$ is decreased.

\cbl
Finally, it is important to recognise that the global response time estimate, Equation (\ref{eq:RTdelta_D}), can also be derived using the classical analytical solution \citep{Carslaw1959} of the homogeneous problem, Equations (\ref{eq:df_model_linear_eq_homogeneous}) and (\ref{eq:df_model_linear_IBCs_homogeneous}). Replacing $h(x,t_r)$ in Equation (\ref{eq:finite_response_time}) with the first term of the Fourier series solution, which is asymptotically equivalent to the complete Fourier series solution as $t\rightarrow\infty$, and solving for the response time, $t_{r}$, yields precisely Equation (\ref{eq:RTdelta_D}). This confirms that in the limit as $k\rightarrow\infty$, our response time estimate (\ref{eq:response_time}) converges precisely to the asymptotic value obtained using the analytical solution. Although one can use the analytical solution for the homogeneous problem, for almost all heterogeneous problems it is not possible to compute the response time in this way as an analytical solution is not available. As a result, one of the main selling points of our new approach for computing the response time is that it works for both homogeneous and heterogeneous problems without the need for the transient (numerical or analytical) solution.
\cb

%which is otherwise unknown.  Here, we find that the constant of proportionality varies with our choice of $\delta$, in a way that makes intuitive sense.  Namely, that the constant of proportionality increases as $\delta$ is decreased.

\section{Results and discussion}
\label{sec:results}
\subsection{Case study for flow in homogeneous porous media}
To illustrate our theoretical developments, we compare the performance of our new method to an experimental data set described previously~\citep{simpson_2013}. In summary, the experiments involve a laboratory-scale aquifer model of width $50$ $\textrm{cm}$ and height $28$ $\textrm{cm}$.  The laboratory-sale aquifer model is packed with uniformly-sized glass beads which we consider to be a homogeneous, isotropic porous medium.  A constant head boundary condition is applied at $x=0\,\mathrm{cm}$, to maintain an initial saturated depth of approximately $18.7\,\mathrm{cm}$.  A no-flow boundary is imposed at the right--hand vertical boundary, where $x=50\,\mathrm{cm}$.  Flow in the laboratory apparatus is initiated through a set of evenly spaced constant flow drippers, installed along the upper boundary of the tank.  We consider two different experiments.  First,  we consider the initial condition in the system to be at a spatially uniform saturated depth $h_0(x) = 18.7$ cm.  At $t=0$, recharge is applied and the increase in saturated thickness at the right hand boundary, where $x=50$ cm, is recorded for $R = 1.23$ cm/min.  Second,  after the recharge has been applied to the laboratory-scale aquifer for a sufficiently long period of time, which we take to be approximately five minutes, the recharge gallery is removed so that the approximately parabolic phreatic surface undergoes a transient discharge response that eventually leads to a horizontal phreatic surface.  We refer to these two transitions as the \textit{recharge} and \textit{discharge} transitions, respectively.

\def\fwidth{0.35\textwidth}
\begin{figure}
\centering
\includegraphics{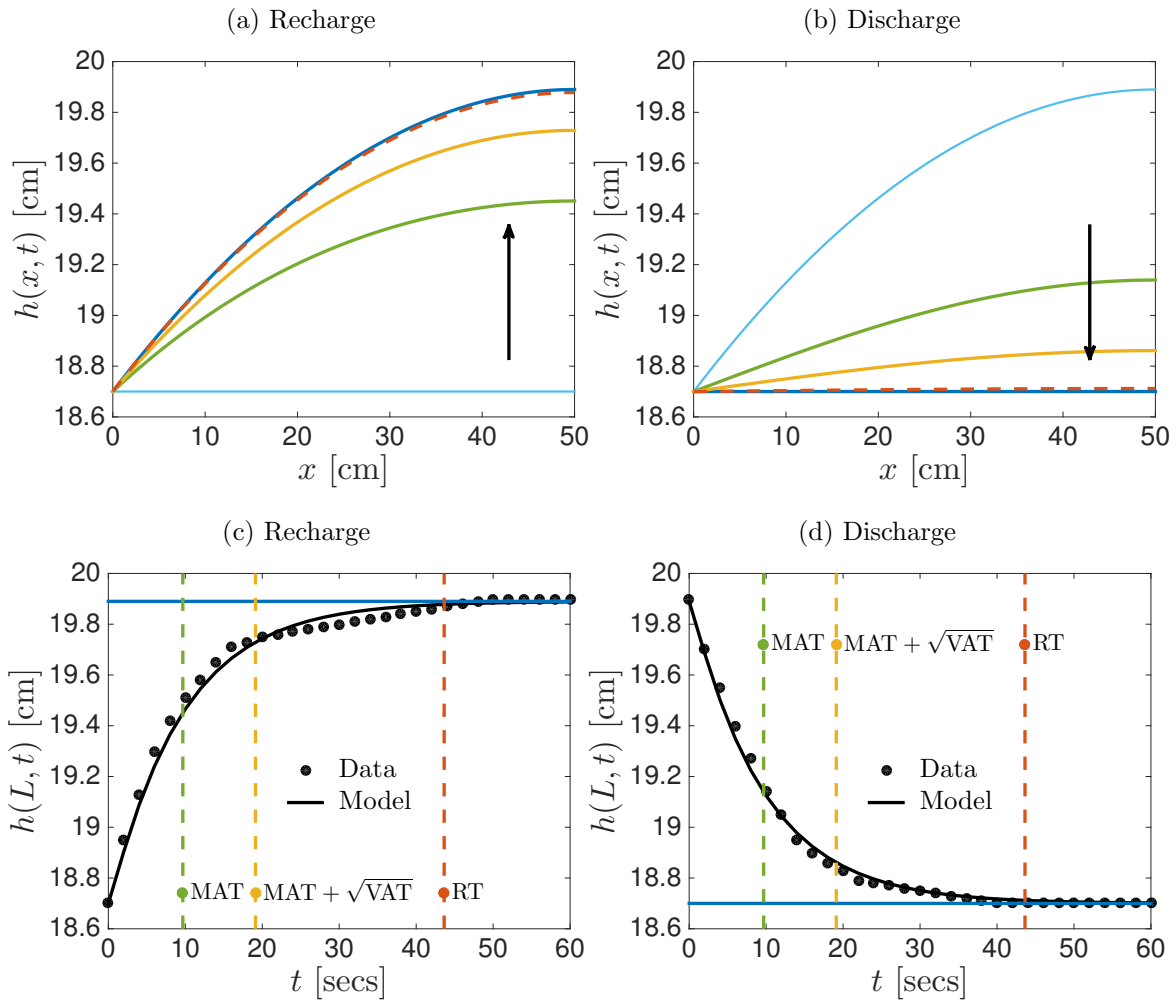}
\caption{\selectfont{}\textbf{Homogeneous flow results.} (a)--(b) Transient solution of Equations (\ref{eq:df_model_linear_eq}) and (\ref{eq:df_model_linear_IBCs}) showing the spatial and temporal variation of the saturated thickness.  Solutions are shown at $t = 0$ (light blue), $t = \mathrm{MAT}(L)$ (green), $t = (\mathrm{MAT}+\sqrt{\mathrm{VAT}})(L)$ (yellow), $t = \mathrm{RT}(L; k,\delta)$ (red dashed) and steady state, as $t \rightarrow\infty$, (dark blue) for (a) the recharge, and (b) the discharge experiments. Arrows indicate the direction of increasing $t$. Profiles in (c)--(d) show the time evolution of the experimental measurements and model predictions of the saturated thickness at $x = L$ (black) for (c) the recharge and (d) the discharge experiments.  The vertical lines are located at $t = \mathrm{MAT}(L)$ (green dashed), $t = (\mathrm{MAT}+\sqrt{\mathrm{VAT}})(L)$ (yellow dashed), $t = \mathrm{RT}(L; k,\delta)$ (red dashed) while the horizontal line denotes $h_{\infty}(L)$ (dark blue). In all figures, $\mathrm{RT}(L; k,\delta)$ is evaluated using $k = 5$ and $\delta = 0.01$. In these experiments we have $\overline{h} = 19\,\mathrm{cm}$, $S = 0.2$, $h_{1} = 18.7\,\mathrm{cm}$ and $L = 50\,\mathrm{cm}$, $K(x) = 68\,\mathrm{cm}/\mathrm{min}$~\citep{simpson_2013}.  The recharge experiment corresponds to $h_{0}(x) = 18.7\,\mathrm{cm}$, $R(x) = 1.23\,\mathrm{cm}/\mathrm{min}$, and the discharge experiment corresponds to $h_{0}(x) = Wx(2L-x)/(2D) + h_{1}$[cm],  $R(x) = 0.0\,\mathrm{cm}/\mathrm{min}$.  In all cases, the spatial discretization of Equations (\ref{eq:df_model_linear_eq}) and (\ref{eq:df_model_linear_IBCs}), Equations (\ref{eq:df_model_linear_eq_ss}) and (\ref{eq:df_model_linear_IBCs_ss}), and Equations (\ref{eq:moments_bvp_ode}) and (\ref{eq:moments_bvp_bcs}) to compute $h(x,t)$, $h_{\infty}(x)$ and $M_{k}(x)$ is carried out using a uniform grid spacing of $50/200 = 0.25\,\mathrm{cm}$ giving $n = 200$ unknown values.}
\label{fig:Homogeneous}
\end{figure}

To illustrate the transition of interest we show, in Figure \ref{fig:Homogeneous}(a) the transient solution, $h(x,t)$, for the homogeneous recharge experiment that has $h_{0}(x) = 18.7\,\mathrm{cm}$ and $R(x) = 1.23\,\mathrm{cm}/\mathrm{min}$.  The numerical estimates of $h(x,t)$ are shown for all $0<x<50\,\mathrm{cm}$, showing that $h(x,t)$ eventually asymptotes to the associated parabolic steady state solution $h_{\infty}(x)$, described in Equation (\ref{eq:steady_homogeneous}), after a sufficiently long period of time.  Similarly, in Figure \ref{fig:Homogeneous}(b), we show the transient solution, $h(x,t)$, for the homogeneous discharge experiment, where the initial parabolic saturated thickness eventually relaxes to $h_{\infty}(x) = 18.7\,\mathrm{cm}$ as $t \to \infty$ since $R(x) = 0\,\mathrm{cm}/\mathrm{min}$ in this case.  Figure \ref{fig:Homogeneous}(c)--(d) also shows experimental data measuring $h(50,t)$ for both transitions.  In both cases we see the asymptotic behaviour as the transient solutions approach the steady state condition.

Results in Figure \ref{fig:Homogeneous}(c) compare the previous low accuracy estimates of the response time with our new asymptotic results.  In particular, the mean action time is approximately 10 seconds, whereas the mean plus one standard deviation is just under 20 seconds for this transition.  Our visual comparison of the rate at which the experimental data for $h(L,t)$ approaches $h_{\infty}(L)$ as $t$ increases implies that there is still a reasonable amount of temporal variation in $h(L,t)$ beyond the amount of time given by $(\mathrm{MAT} + \sqrt{\mathrm{VAT}})(L)$. In comparison, calculating our new estimate of the response time $\mathrm{RT}(L; k,\delta)$, by setting $k=5$ and $\delta =0.01$, gives a value of approximately 44 seconds which, when plotted at the scale in Figure \ref{fig:Homogeneous}(c), accurately captures the smallest value of $t$ at which $h(L,t)$ and $h_{\infty}(L)$ are visually indistinguishable. Therefore, the new estimate of the response time provides a far more accurate estimate than previous estimates based on the theory of mean action time~\citep{simpson_2013,jazaei_2014}.  Furthermore, our interpretation of the accuracy of $\mathrm{MAT}(L)$, $(\mathrm{MAT} + \sqrt{\mathrm{VAT}})(L)$ and $\mathrm{RT}(L; 5, 0.01)$ for the recharge experiment in Figure \ref{fig:Homogeneous}(c), also applies to the estimates in Figure \ref{fig:Homogeneous}(d) for the discharge experiments.  As before, the mean action time is approximately 10 seconds, and the mean plus one standard deviation is just under 20 seconds, and our visual interpretation of the data suggests that significant temporal variations occur beyond these durations of time.  In contrast, the new estimate is approximately 44 seconds which accurately captures the smallest value of $t$ at which $h(L,t)$ and $h_{\infty}(L)$ are visually indistinguishable, again confirming that the new estimate is far more accurate than previous estimates.

\begin{table}
\centering
\small
\begin{tabular}{lrrrrrr}
\hline
\setlength{\tabcolsep}{5pt}
& \multicolumn{3}{c}{Recharge} & \multicolumn{3}{c}{Discharge}\\
& $t_{r}$ & $\delta_{r}$ & $|\delta_{r} - \delta|$ & $t_{r}$ & $\delta_{r}$ & $|\delta_{r} - \delta|$\\
\cline{2-7}
$\mathrm{MAT}(L)$ & 9.6751 & 0.37 & 3.6e-01 & 9.6751 & 0.37 & 3.6e-01\\
$(\mathrm{MAT}+\sqrt{\mathrm{VAT}})(L)$ & 19.1152 & 0.14 & 1.3e-01 & 19.1152 & 0.14 & 1.3e-01\\
$\mathrm{RT}(L; 1,\delta)$ & 44.5556 & 0.01 & 9.3e-04 & 44.5556 & 0.01 & 9.3e-04\\
$\mathrm{RT}(L; 2,\delta)$ & \cellcolor{blue!15}43.7157 & 0.01 & 8.5e-05 & \cellcolor{blue!15}43.7157 & 0.01 & 8.5e-05\\
$\mathrm{RT}(L; 3,\delta)$ & 43.6410 & 0.01 & 6.0e-06 & 43.6410 & 0.01 & 6.0e-06\\
$\mathrm{RT}(L; 4,\delta)$ & 43.6356 & 0.01 & 2.3e-07 & 43.6356 & 0.01 & 2.3e-07\\
$\mathrm{RT}(L; 5,\delta)$ & 43.6353 & 0.01 & 2.4e-08 & 43.6353 & 0.01 & 2.4e-08\\
$\mathrm{RT}(L; 6,\delta)$ & \cellcolor{red!15}43.6354 & 0.01 & 8.3e-09 & \cellcolor{red!15}43.6354 & 0.01 & 8.3e-09\\
$\mathrm{RT}(L; 7,\delta)$ & 43.6354 & 0.01 & 1.5e-09 & 43.6354 & 0.01 & 1.5e-09\\
$\mathrm{RT}(L; 8,\delta)$ & 43.6354 & 0.01 & 2.4e-10 & 43.6354 & 0.01 & 2.4e-10\\
$\mathrm{RT}(L; 9,\delta)$ & 43.6354 & 0.01 & 3.4e-11 & 43.6354 & 0.01 & 3.4e-11\\
$\mathrm{RT}(L; 10,\delta)$ & 43.6354 & 0.01 & 4.7e-12 & 43.6354 & 0.01 & 4.8e-12\\
\hline
\end{tabular}
\caption{\selectfont{}\textbf{Homogeneous flow results.} Response time estimates for the homogeneous recharge and discharge experiments with a prescribed tolerance of $\delta = 0.01$, where $t_{r}$ denotes the chosen response time estimate and $\delta_{r} = 1 - F(t_{r}; L) = (h(L,t_{r}) - h_{\infty}(L))/(h_{0}(L) - h_{\infty}(L))$.   Note that $\delta_{r} = \delta$ when $t_{r}$ is exact. Included in the table are the following quantities all evaluated at $x = L$ where the maximum response time occurs: the mean action time $\mathrm{MAT}(x)$, given by Equation (\ref{eq:MAT}); the mean plus one standard deviation of action time $(\mathrm{MAT}+\sqrt{\mathrm{VAT}})(x)$, given by Equaiton (\ref{eq:MATVAT}); and the new estimate of the response time  $\mathrm{RT}(x; k,\delta)$, given by Equation (\ref{eq:response_time}) for $k = 1,2,\hdots,10$.  In all cases, the spatial discretization of Equations (\ref{eq:df_model_linear_eq}) and (\ref{eq:df_model_linear_IBCs}), Equations (\ref{eq:df_model_linear_eq_ss}) and (\ref{eq:df_model_linear_IBCs_ss}), and Equations (\ref{eq:moments_bvp_ode}) and (\ref{eq:moments_bvp_bcs}) to compute $h(x,t)$, $h_{\infty}(x)$ and $M_{k}(x)$ is carried out using a uniform grid spacing of $50/200 = 0.25\,\mathrm{cm}$ giving $n = 200$ unknown values. The blue-shaded cells indicate when the response time is correct to the nearest second, while the red-shaded cells indicate when the response time has converged at the fourth decimal place.}
\label{tab:Homogeneous}
\end{table}

We now examine the question of how many moments do we need to calculate for our asymptotic results to be practically useful.  To explore this question we present, in Table \ref{tab:Homogeneous}, a comparison of $\mathrm{RT}(L; k, \delta)$ for $k=1,2,3, \dots, 10$ and $\delta = 0.01$ for both the recharge and discharge problems.  There are several interesting results.  First, all estimates of the response time are equivalent for the recharge and discharge problem, which makes intuitive sense since the discharge problem is just the recharge problem in reverse. Second, our estimates of the response time $\mathrm{RT}(L; k,\delta)$ converge to a highly accurate asymptotic approximation of the response time $t_{r}(L)$, as defined in Equation (\ref{eq:finite_response_time}), as the value of $\delta_{r} = (h(L,t_{r}) - h_{\infty}(L))/(h_{0}(L) - h_{\infty}(L))$ evaluated at $t_{r}=\mathrm{RT}(L; k,\delta)$ is approaching $\delta = 0.01$ as $k$ increases. Third, the larger values of $\delta_{r}$ at $t_{r} = \mathrm{MAT}(L)$ and $t_{r} = (\mathrm{MAT}+\sqrt{\mathrm{VAT}})(L)$ confirm the lower accuracy of these estimates. Fourth, our estimates of the response time converge, at the fourth decimal place, by $k=6$ (red-shaded cells in Table \ref{tab:Homogeneous}). \cbl Finally, if we are satisfied with estimating the response time to the nearest second for this problem then $k = 2$ is sufficient (blue-shaded cells in Table \ref{tab:Homogeneous}), which means we need only compute the first two moments, $M_{1}(x)$ and $M_{2}(x)$. These results are extremely encouraging.  Not only does our approach for calculating the response time avoid solving the underlying transient flow model, implementing these results for a practical problem shows that dealing with just $k=2$ moments is sufficient for practical purposes. \cb
 %, but we also make use of asymptotic results that are, strictly speaking, valid for sufficiently large $k$.  Yet,

\subsection{Case study for flow in heterogeneous porous media}
We now demonstrate how the new method for computing response times applies to the more practical case where flow takes place in a heterogeneous porous medium. To examine this we maintain the same geometry, initial conditions, boundary conditions and some of the material properties from the first case study, namely we consider a recharge problem with a uniform initial condition, $h_{0}(x) = 18.7\,\mathrm{cm}$, recharge rate of $R(x) = 1.23\,\mathrm{cm}/\mathrm{min}$, an average saturated thickness of $\overline{h} = 19\,\mathrm{cm}$, and a storage coefficient of $S = 0.2$. The key difference is that now we consider three different kinds of spatial variations in the saturated hydraulic conductivity,
\begin{itemize}
\item Case A: $K(x) = 83.4879 - 64\exp(-0.1(x-25)^2)$
\item Case B: $K(x) = 63.3598 + 64\exp(-0.1(x-25)^2)$
\item Case C: $K(x) = 81.0707 + 64\exp(-0.1(x-50/3)^2)-64\exp(-0.1(x-100/3)^2)$.
\end{itemize}

\begin{figure}
\includegraphics{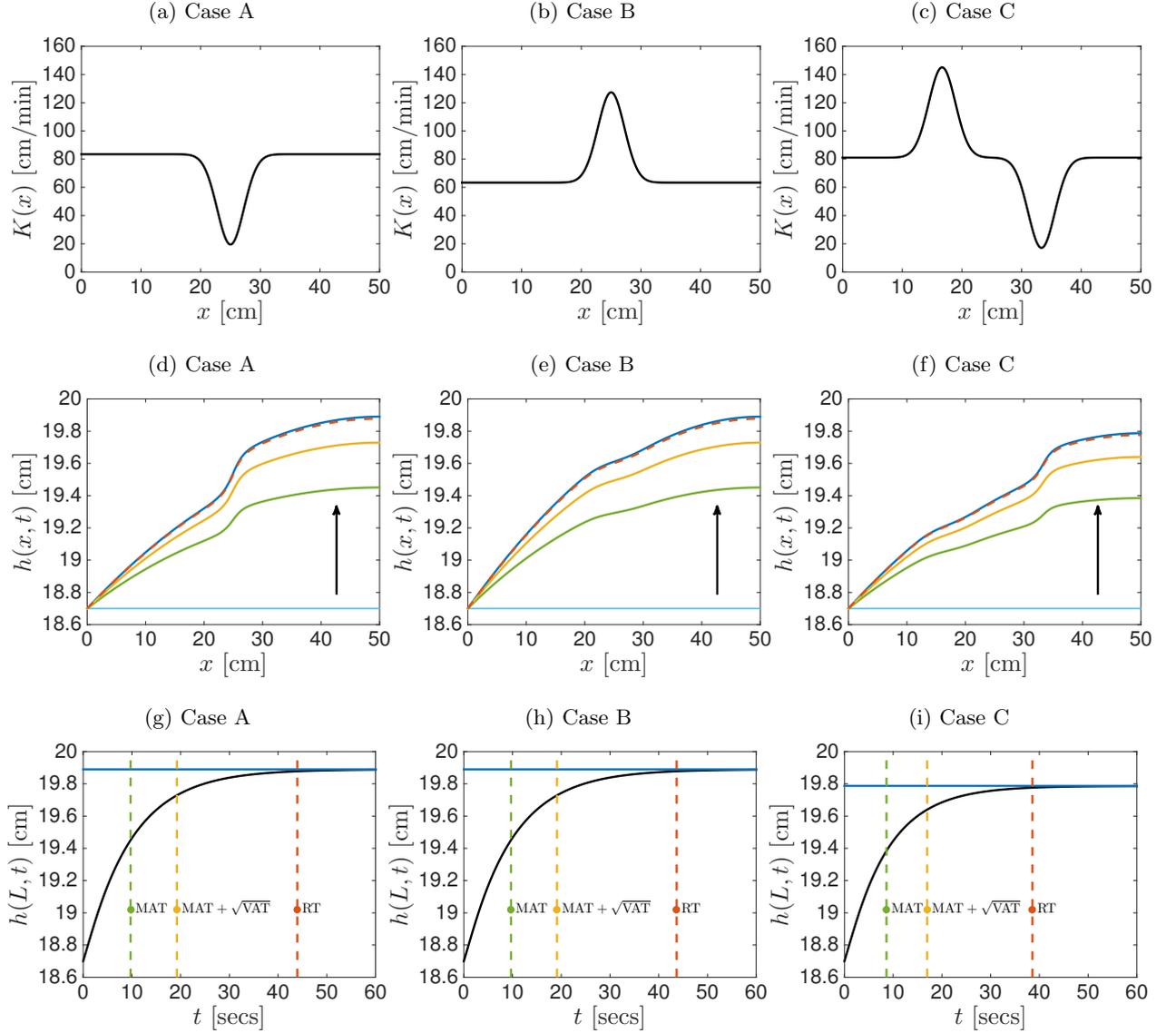}
\caption{\selectfont{}\textbf{Heterogeneous flow results.} (a)--(c) Spatially-dependent saturated hydraulic conductivity functions for Cases A--C, respectively (d)--(f) Transient solution of the groundwater flow model, Equations (\ref{eq:df_model_linear_eq}) and (\ref{eq:df_model_linear_IBCs}), depicting the spatial variation of the saturated thickness at $t = 0$ (light blue), $t = \mathrm{MAT}(L)$ (green), $t = (\mathrm{MAT}+\sqrt{\mathrm{VAT}})(L)$ (yellow), $t = \mathrm{RT}(L; k,\delta)$ (red dashed) and steady state ($t \rightarrow\infty$) (dark blue) for Cases A--C, respectively, with arrows indicating the direction of increasing time. (g)-(i) Evolution of the saturated thickness at the left boundary ($x = L$) over time (black) for Cases A--C, respectively. The vertical lines are located $t = \mathrm{MAT}(L)$ (green dashed), $t = (\mathrm{MAT}+\sqrt{\mathrm{VAT}})(L)$ (yellow dashed), $t = \mathrm{RT}(L; k,\delta)$ (red dashed) while the horizontal line denotes the steady state value of the saturated thickness at the left boundary ($x = L$) (dark blue). In each figure (d)--(i), $\mathrm{RT}(L; k,\delta)$ is evaluated using $k = 5$ and $\delta = 0.01$. In all cases, the spatial discretization of Equations (\ref{eq:df_model_linear_eq}) and (\ref{eq:df_model_linear_IBCs}), Equations (\ref{eq:df_model_linear_eq_ss}) and (\ref{eq:df_model_linear_IBCs_ss}), and Equations (\ref{eq:moments_bvp_ode}) and (\ref{eq:moments_bvp_bcs}) to compute $h(x,t)$, $h_{\infty}(x)$ and $M_{k}(x)$ is carried out using a uniform grid spacing of $50/200 = 0.25\,\mathrm{cm}$ giving $n = 200$ unknown values.}
\label{fig:Heterogeneous}
\end{figure}

Case A corresponds to a constant background saturated hydraulic conductivity with a localised region of increased conductivity centred at $x=25\,\mathrm{cm}$, whereas case B considers a constant background saturated hydraulic conductivity with a localised region of decreased conductivity at $x=25\,\mathrm{cm}$.  Case C is a more complicated case that contains both a localised region of increased conductivity and decreased conductivity, at location $x=50/3\,\mathrm{cm}$ and $x=100/3\,\mathrm{cm}$, respectively.  Plots of $K(x)$ for these cases are given in Figure \ref{fig:Heterogeneous}(a)-(c).  The particular constants appearing in these $K(x)$ functions are rounded to four decimal places and chosen so that the harmonic average of the saturated hydraulic conductivity functions are equivalent, ${\overline{K} = L/ (\int_{0}^{L} 1/K(x)\,\textrm{d}x) = 68\,\mathrm{cm}/\mathrm{min}}$.  Therefore, the hydraulic conductivity functions lead to the same average conductivity as we examined previously in the homogeneous case.  Since we do not have laboratory data for these three cases of flow in heterogeneous porous media, we solve Equations (\ref{eq:df_model_linear_eq}) and (\ref{eq:df_model_linear_IBCs}) numerically to obtain $h(x,t)$ for each recharge experiment with the three different $K(x)$ functions.  Results are summarised in Figure \ref{fig:Heterogeneous}. %Furthermore, once we have examined the response time for the recharge experiment for each of the three $K(x)$ functions, we then set $R(x)=0$ and examine the corresponding discharge transition.

Results in Figure \ref{fig:Heterogeneous}(d)-(f) show the temporal evolution of $h(x,t)$ for the recharge transition for these three different forms of spatial heterogeneity in $K(x)$.  Comparing the three different sets of results confirms that the shape of the $h(x,t)$ profiles depends on the form of $K(x)$, and the impact of spatial variations in $K(x)$ become pronounced at later times as $h(x,t)$ tends to approach the corresponding steady state profile, $h_{\infty}(x)$.  Results in Figure \ref{fig:Heterogeneous}(g)-(i) compare three estimates of the transition time with the numerical solution showing how  $h(L,t)$ asymptotes to  $h_{\infty}(L)$.  For all three forms of spatial heterogeneity we see that the mean action time is approximately 10 seconds for cases A and B, and 9 seconds for case C, yet a visual comparison of the numerical solutions showing $h(L,t)$ indicates that there is still a relatively large transient response after this time.  Similarly, the quantity $(\textrm{MAT} + \sqrt{\textrm{VAT}})(L)$ is also plotted and we also see a modest transient response after this time.  In contrast, our estimate of the response time, $\mathrm{RT}(L; 5, \delta)$, accurately captures the smallest value of $t$ at which $h(L,t)$ and $h_{\infty}(L)$ are visually indistinguishable at the scale shown in Figures \ref{fig:Heterogeneous}(g)--(i).  Therefore, our new method provides very accurate estimates of the response time for flows in heterogeneous porous media.

\begin{table}[h]
\centering
\setlength{\tabcolsep}{5pt}
\small
\begin{tabular}{lrrrrrrrrr}
\hline
& \multicolumn{3}{c}{Case A} & \multicolumn{3}{c}{Case B} & \multicolumn{3}{c}{Case C}\\
& $t_{r}$ & $\delta_{r}$ & $|\delta_{r} - \delta|$ & $t_{r}$ & $\delta_{r}$ & $|\delta_{r} - \delta|$ & $t_{r}$ & $\delta_{r}$ & $|\delta_{r} - \delta|$\\
\cline{2-10}
$\mathrm{MAT}(L)$ & 9.7278 & 0.37 & 3.6e-01 & 9.6823 & 0.37 & 3.6e-01 & 8.6281 & 0.37 & 3.6e-01\\
$(\mathrm{MAT}+\sqrt{\mathrm{VAT}})(L)$ & 19.2329 & 0.14 & 1.3e-01 & 19.1311 & 0.14 & 1.3e-01 & 16.9670 & 0.14 & 1.3e-01\\
$\mathrm{RT}(L; 1,\delta)$ & 44.7983 & 0.01 & 8.8e-04 & 44.5884 & 0.01 & 9.2e-04 & 39.7338 & 0.01 & 1.3e-03\\
$\mathrm{RT}(L; 2,\delta)$ & \cellcolor{blue!15}44.0021 & 0.01 & 7.8e-05 & \cellcolor{blue!15}43.7543 & 0.01 & 8.4e-05 & \cellcolor{blue!15}38.7041 & 0.01 & 1.6e-04\\
$\mathrm{RT}(L; 3,\delta)$ & 43.9332 & 0.01 & 5.3e-06 & 43.6803 & 0.01 & 5.9e-06 & 38.5858 & 0.01 & 1.5e-05\\
$\mathrm{RT}(L; 4,\delta)$ & 43.9284 & 0.01 & 2.0e-07 & 43.6750 & 0.01 & 2.2e-07 & 38.5743 & 0.01 & 7.2e-07\\
$\mathrm{RT}(L; 5,\delta)$ & \cellcolor{red!15}43.9282 & 0.01 & 2.1e-08 & \cellcolor{red!15}43.6748 & 0.01 & 2.4e-08 & 38.5736 & 0.01 & 1.3e-07\\
$\mathrm{RT}(L; 6,\delta)$ & 43.9282 & 0.01 & 6.9e-09 & 43.6748 & 0.01 & 8.1e-09 & \cellcolor{red!15}38.5737 & 0.01 & 5.6e-08\\
$\mathrm{RT}(L; 7,\delta)$ & 43.9282 & 0.01 & 1.3e-09 & 43.6748 & 0.01 & 1.5e-09 & 38.5737 & 0.01 & 1.4e-08\\
$\mathrm{RT}(L; 8,\delta)$ & 43.9282 & 0.01 & 1.9e-10 & 43.6748 & 0.01 & 2.3e-10 & 38.5737 & 0.01 & 3.0e-09\\
$\mathrm{RT}(L; 9,\delta)$ & 43.9282 & 0.01 & 2.7e-11 & 43.6748 & 0.01 & 3.3e-11 & 38.5737 & 0.01 & 6.0e-10\\
$\mathrm{RT}(L; 10,\delta)$ & 43.9282 & 0.01 & 3.6e-12 & 43.6748 & 0.01 & 4.5e-12 & 38.5737 & 0.01 & 1.1e-10\\
\hline
\end{tabular}
\caption{\selectfont{}\textbf{Heterogeneous flow results.} Response time estimates for the heterogeneous test cases with a prescribed tolerance of $\delta = 0.01$, where $t_{r}$ denotes the chosen response time estimate and $\delta_{r} = 1 - F(t_{r}; L) = (h(L,t_{r}) - h_{\infty}(L))/(h_{0}(L) - h_{\infty}(L))$.  Note that $\delta_{r} = \delta$ when $t_{r}$ is exact. Included in the table are the following quantities all evaluated at $x = L$ where the maximum response time occurs: the mean action time $\mathrm{MAT}(x)$, given by Equation (\ref{eq:MAT}); the mean plus one standard deviation of action time $(\mathrm{MAT}+\sqrt{\mathrm{VAT}})(x)$, given by Equation (\ref{eq:MATVAT}); and the new estimate of the response time  $\mathrm{RT}(x; k,\delta)$, given by Equation (\ref{eq:response_time}) for $k = 1,2,\hdots,10$. In all cases, the spatial discretization of Equations (\ref{eq:df_model_linear_eq}) and (\ref{eq:df_model_linear_IBCs}), Equations (\ref{eq:df_model_linear_eq_ss}) and (\ref{eq:df_model_linear_IBCs_ss}), and Equations (\ref{eq:moments_bvp_ode}) and (\ref{eq:moments_bvp_bcs}) to compute $h(x,t)$, $h_{\infty}(x)$ and $M_{k}(x)$ is carried out using a uniform grid spacing of $50/200 = 0.25\,\mathrm{cm}$ giving $n = 200$ unknown values. The blue-shaded cells indicate when the response time is correct to the nearest second, while the red-shaded cells indicate when the response time has converged at the fourth decimal place.}
\label{tab:Heterogeneous}
\end{table}

In Table \ref{tab:Heterogeneous}, computed values of $\mathrm{RT}(L; k,\delta)$ are listed for $k = 1,2,3,\hdots,10$ and $\delta = 0.01$ for cases A--C. Several interesting observations are evident. The estimate of the response time is 44 seconds for cases A and B and 39 seconds for case C when rounded to the nearest second. These results demonstrate that it is not always sufficient to calculate the response time based on the average value of the saturated hydraulic conductivity. In fact, substituting $K = \overline{K} = 68\,\mathrm{cm}/\mathrm{min}$ into Equation (\ref{eq:RTdelta_D}) yields a response time of 44 seconds when rounded to the nearest second, which overestimates the response time for case C by approximately 5 seconds. As observed for the homogeneous test cases, our estimates of the response time $\mathrm{RT}(L; k,\delta)$ converge to a highly accurate asymptotic approximation of the response time $t_{r}(L)$, as defined in Equation (\ref{eq:finite_response_time}), since the value of $\delta_{r} = (h(L,t_{r}) - h_{\infty}(L))/(h_{0}(L) - h_{\infty}(L))$ evaluated at $t_{r}=\mathrm{RT}(L; k,\delta)$ is approaching $\delta = 0.01$ as $k$ increases. Again, the larger values of $\delta_{r}$ at $t_{r} = \mathrm{MAT}(L)$ and $t_{r} = (\mathrm{MAT}+\sqrt{\mathrm{VAT}})(L)$ confirm the lower accuracy of these estimates compared to our new estimate. To achieve four decimal places of accuracy, $k = 5$ (cases A and B) and $k = 6$ (case C) is sufficient (red-shaded cells in Table \ref{tab:Heterogeneous}) while an estimate of the response time, accurate to the nearest second, is obtained using our new estimate with only the first two moments ($k = 2$, blue-shaded cells in Table \ref{tab:Heterogeneous}), as was observed for the homogeneous recharge and discharge problems.

\section{Summary and conclusions}
\label{sec:conclusions}
In this work, we present a new method for calculating highly accurate estimates of response times for groundwater flow processes. The analysis is carried out using the linearised one-dimensional Dupuit-Forchheimer model of saturated flow through a heterogeneous porous medium. Our strategy is to calculate the time required to effectively reach steady state, defined as the time at which the proportion of the transient process remaining is equal to a small specified tolerance $\delta$. Our approach extends the concept of mean action time by computing higher-order raw moments of the probability density function associated with the transition from the initial condition to the steady state condition. By studying the long time behaviour of the corresponding cumulative distribution function, we derive a simple formula for the response time depending on the specified tolerance $\delta$ and two consecutive raw moments, $k-1$ and $k$, for a suitable choice of $k$. Attractively, the new approach does not require the solution of the transient groundwater flow problem.

Our new estimate of the response time is significantly more accurate than existing estimates that are based on the first two central moments only. This is demonstrated by presenting two case studies, the first involving flow in homogeneous media and a comparison to a suite of laboratory-scale experiments, and the second involving flow in heterogeneous media. Our new estimate of the response time converges to a highly accurate asymptotic approximation of the response time as $k$ increases and is able to accurately capture the response time evident in the experimental data for the homogeneous problem. Across both case studies, setting $k = 2$ produces an estimate of the response time (measured in seconds) correct to the nearest second while setting $k=6$ produces an estimate accurate to four decimal places. In comparison, existing estimates based on the first two central moments are shown to significantly underestimate the response time as evident by the amount of temporal variation that takes place beyond these points in time.

Our new approach requires significantly less computational effort than the existing approach of studying the response time using the transient solution of the groundwater flow model. Using standard spatial and temporal discretization methods, the computational cost of both approaches is dominated by the solution of linear systems of size $n \times n$, where $n$ denotes the number of discrete unknown values utilised in the spatial discretization. Our new approach requires the solution of $m+1$ linear systems of size $n\times n$ for $k = m$ while using the transient solution to study the response time, requires the solution of a linear system of size $n\times n$ at each time step. As mentioned above, very accurate estimates are obtained using the new approach for $m = \text{3, 4 or 5}$, which means that our new approach requires the solution of between four and six linear systems. Comparatively, the number of time steps required in a typical transient simulation is at least an order of magnitude greater this number, and possibly several orders of magnitude greater. Hence, using the transient solution to study the response time is significantly more computationally expensive as one requires the solution of many more linear systems of size $n\times n$.

For a specific problem of flow in homogeneous porous media, we utilise our new method to derive a very simple result that demonstrates that the response time takes the form of a constant, depending on the specified tolerance $\delta$, multiplied by the time scale $L^{2}/D$, where $L$ is the length of the aquifer and $D$ is the aquifer diffusivity. This analysis provides a rigorous mathematical connection with the often used scaling argument that states that the response time is proportional to $L^{2}/D$. Moreover, we explicitly give the constant of proportionality for several common choices of the specified tolerance $\delta$.

\change{For some problems}, employing an absolute measure to determine how close the transient solution is to steady state \change{is} preferred over the relative measure, Equation (\ref{eq:finite_response_time}), used in the analysis presented in this paper. To specify an absolute tolerance, the response time, $t = t_{r}$, is instead defined to satisfy $h(x,t_r) - h_{\infty}(x) = \delta$ rather than Equation (\ref{eq:finite_response_time})\footnote{Note that according to this new definition, one must specify a negative value for the absolute tolerance, $\delta$, if the transient solution, $h(x,t)$, increases from $h_{0}(x)$ to $h_{\infty}(x)$.}. Importantly, reformulating our approach for calculating the response time in terms of an absolute tolerance does not introduce any complications. Simply replacing $\delta$ in the response time formula, Equation (\ref{eq:response_time}), by $\delta/(h_{0}(x) - h_{\infty}(x))$  yields the modified response time formula for a specified absolute tolerance $\delta$:
\begin{align*}
\mathrm{RT}(x; k,\delta) = \frac{M_{k}(x)}{kM_{k-1}(x)}\log_{\textrm{e}}\left[\frac{M_{k}(x)(h_{0}(x)-h_{\infty}(x))}{k! \,\delta}\left(\frac{kM_{k-1}(x)}{M_{k}(x)}\right)^{k}\right].
\end{align*}

This paper focusses on groundwater flow processes through a one-dimensional heterogeneous porous medium. While the analysis and results are presented for one-dimensional flow only, the techniques presented carry over to two and three dimensional problems with the raw moments satisfying two and three dimensional boundary value problems, respectively. Furthermore, due to the increased size of the linear systems arising from spatial discretization in higher dimensions, we expect that the computational savings will be even more pronounced than those reported here for the one-dimensional problem. Another simplification in this study is that we consider transient flow conditions that are established by instantaneous forcing conditions by, for example, the sudden application or removal of recharge.  In practice, such transitions might be driven by a more subtle time-dependent forcing condition and we note that our analysis can be extended to time-dependent forcing conditions by adopting the approach of \cite{jazaei_2014}.

\change{A final comment is that the focus of this study is on the development of mathematical expressions to calculate the response time.  In this work, we calculate the response time without solving the underlying transient groundwater flow equation for $h(x,t)$. However, it is also worthwhile pointing out that our method can be used directly with field data if measurements of the saturated thickness, $h(x,t)$, are available as a time series. In this case, quadrature can be used to calculate the raw moments from the field data according to Equations (\ref{eq:pdf}) and (\ref{eq:raw_moment}).  Given these computed moments, one can then apply the response time formula, Equation (\ref{eq:response_time}), to estimate the response time.}

\section*{Acknowledgements} This work is supported by the Australian Research Council (DE150101137, DP170100474). We thank Willem Zaadnoordijk and the two anonymous referees for their helpful comments.

%\section*{References}
\bibliographystyle{plainnat}
\bibliography{references}

\end{document}